\documentclass{article}
\usepackage{amsfonts}

\newsymbol{\lesssim} 132E 
\newsymbol{\gtrsim}  1326 

\newcommand{\opname}[1]{\mathop{\rm#1}\nolimits} 

\newcommand{\Aslash}{{A\mkern-9mu/}} 
\newcommand{\braket}[2]{\langle#1\mathbin\vert#2\rangle} 
\newcommand{\delslash}{{\partial\mkern-9mu/}} 
\newcommand{\GeV}{\opname{GeV}}    
\newcommand{\mtop}{m_{\rm top}}    
\newcommand{\pbraket}[2]{(#1\mathbin\vert#2)} 
\newcommand{\set}[1]{\{\,#1\,\}}   
\newcommand{\Str}{\opname{Str}}    
\newcommand{\tfrac}[2]{{\textstyle{#1\over#2}}} 
\newcommand{\tr}{\opname{tr}}      

\newbox\ncgintdbox \newbox\ncginttbox
\setbox0=\hbox{$-$}
\setbox2=\hbox{$\displaystyle\int$}
\setbox\ncgintdbox=\hbox{\rlap{\hbox
    to \wd2{\hskip-.125em\box2\relax\hfil}}\box0\kern.1em}
\setbox0=\hbox{$\vcenter{\hrule width 4pt}$}
\setbox2=\hbox{$\textstyle\int$}
\setbox\ncginttbox=\hbox{\rlap{\hbox
    to \wd2{\hskip-.125em\box2\relax\hfil}}\box0\kern.1em}
\def\ncgint{\mathop{\mathchoice{\copy\ncgintdbox}{\copy\ncginttbox}
                     {\copy\ncginttbox}{\copy\ncginttbox}}\nolimits}

\title{Noncommutative Geometry and the\\
       Standard Model: An overview}

\author{Jos\'e M. Gracia-Bond{\'\i}a \\
        Department of Mathematics, Universidad de Costa Rica,\\
        2060 San Jos\'e, Costa Rica}
\date{}

\begin{document}

\maketitle

\subsection*{Introduction}

As it is usually presented in textbooks, the Standard Model of
``fundamental interactions'' is, mathematically speaking, a hideous
construction. We can summarize its content thus:

\paragraph{1}
The interaction fields are gauge fields with a
$SU(3)_c \times SU(2)_L \times U(1)_Y$ symmetry. In other words, there
are three systems of gauge bosons $A_\nu^{SU(3)}$, $A_\nu^{SU(2)}$,
$A_\nu^{U(1)}$. They contribute terms
$\pbraket{F}{F} := \tr \int F_{\mu\nu} F^{\mu\nu} \,d^4x$ to the
Action, where $F$ is the gauge field, which is obtained from the gauge
potential by the recipe $F = dA + A \wedge A$.

\paragraph{2}
The (fermionic) matter fields $\psi$ contribute terms
$\int \psi_1 D \psi_2
 = \int \psi_1 \delslash \psi_2 + \int \psi_1 \Aslash \psi_2$.

\paragraph{3}
Unfortunately, in order to give mass to the electroweak gauge bosons,
there is the need to add a colorless scalar ``matter'' field, called
the Higgs particle, with dynamics given by
$\int D^{\nu}\phi^\dagger \,D_{\nu}\phi + V(\phi)$ where
$V(\phi) = -\mu^2\phi^\dagger \phi + \lambda(\phi^\dagger \phi)^2$.
The ``negative mass'' $\mu$ is needed for symmetry breakdown to work.
The introduction of the Higgs is justified on a technical basis: it
preserves unitarity and renormalizability of the quantized theory and
\dots\ it works. It also gives mass to the fermions through the
seemingly ad~hoc and apparently non gauged \dots

\paragraph{4}  \dots\ Yukawa interaction terms,
$\int \bar\psi_1 \phi \psi_2$.

\paragraph{5}
We summarize thus the several aesthetically unpleasant features of the
SM:

\begin{enumerate}
\item
The Higgs sector is introduced by hand.
\item
The link between the parity violating and the symmetry breaking sector
remains mysterious.
\item
There is no explanation for the observed number of fermionic
generations.
\item
The choice of gauge groups and hypercharge assignments seems rather
arbitrary, although it has the felicitous result that the model,
despite being chiral, is anomaly-free.
\item
There is an apparent juxtaposition of gauged and non-gauged interaction
sectors.
\item
There is no explanation for the huge span of fermionic masses.
\end{enumerate}

Noncommutative geometry goes a good bit of the way to solving these
questions ---except the last.

\subsection*{A new framework for thinking about the SM}

In noncommutative geometry (NGC) all the complexities and
idiosyncrasies of the SM stem from a ``pure QCD-like theory'' with a
unified noncommutative gauge boson $\Bbb A$ for the
$SU(3)_c \times SU(2)_L \times U(1)_Y$ symmetry. Thus the Lagrangian:
$$
{\cal L}_{\rm NCG} = -\tfrac14 \pbraket{\Bbb F}{\Bbb F}
                     + \braket{\bar\Psi}{D({\Bbb A})\Psi}
$$
on a {\it noncommutative space\/}, to wit, the {\it product of $M_4$ by
the space of the internal degrees of freedom\/}: colour, weak isospin
and hypercharge. Here
$$
{\Bbb A} = {\Bbb A}(A^{SU(3)}, A^{SU(2)}, A^{U(1)}, \phi).
$$
That is to say, the Higgs is seen as a gauge boson (this helps to
explain its quartic kinetic energy and its pointlike coupling to
fermions). We still have ${\Bbb F} = d{\Bbb A} + {\Bbb A}^2$, and
therefore
$$
{\Bbb F} = {\Bbb F}(F^{SU(3)}, F^{SU(2)}, F^{U(1)}, D\phi, |V|^{1/2}).
$$

\subsection*{The spaces of noncommutative geometry}

The mathematical framework hinges on two related ideas: (1) geometrical
properties of spaces of points (e.g., spacetime without chirality) are
determined by their $c$-number functions; (2) other geometrical settings
(e.g., spacetime with chirality) can be accommodated by allowing
noncommutative algebras of $q$-number functions; both are thought of as
algebras of operators on Hilbert spaces.

Many structures arising in classical geometry are thus replaced by their
quantum counterparts. For instances, measure spaces are replaced by
von~Neu\-mann algebras, topological spaces by $C^*$-algebras, vector
bundles by projective modules, Lie groups by smooth groupoids, de~Rham
homology by cyclic cohomology, and spin manifolds by spectral triples.

Think of functions as forming an algebra $\cal A$ of multiplication
operators on a Hilbert space ${\cal H} = {\cal H}^+ \oplus {\cal H}^-$.
If $\Gamma$ is the sign operator (${} = \pm 1$ on ${\cal H}^\pm$), then
$\delta f = [\Gamma,f]$ is an ``infinitesimal'' operator.  Differential
calculus is done with a ``spectral triple'' consisting of the algebra
$\cal A$, the Hilbert space $\cal H$ and an odd selfadjoint operator $D$
on~$\cal H$ (e.g., the Dirac operator on the space of spinors
$L^2(S_M)$).  Integration of functions is effected by the Dixmier trace
of operators: if $T$ has eigenvalues $\mu_n(T) \geq 0$, then
$$
\ncgint T = \lim_{n\to\infty} \frac{\mu_0(T) +\cdots+ \mu_n(T)}{\log n},
  \qquad\mbox{where}\quad   \int f = \ncgint f |D|^{-d}.
$$

Other classical geometrical objects have their quantum counterparts.
A complex variable becomes an operator in $\cal H$, a real variable is
a selfadjoint operator, and an infinitesimal is a compact operator. An
infinitesimal of order $k$ is seen to be a compact operator whose
singular values $\mu_n$ are $O(n^{-k})$ as $n\to\infty$. The
differential of real or complex variable is replaced by
$\delta f \equiv [\Gamma,f] = \Gamma f - f\Gamma$; and the integral of
a first-order infinitesimal is given by the Dixmier trace.

The spectral triple $({\cal A},{\cal H},D)$ determines the geometry
completely. For example, here is the formula for computing distances
between points (i.e., pure states of $\cal A$) on a conventional
Riemannian manifold:
$$
d(p,q) = \sup\set{ |f(p) - f(q)| : f \in {\cal A},\ \|[D,f]\| \leq 1},
$$
where $D = \delslash$ is the usual Dirac operator. Thus, we now have a
fully quantum formalism for the classical world, and we notice that
distances are better measured by neutrinos than by scalar particles!

\subsection*{The reconstruction of the SM}

We need to have more details on the noncommutative differential
calculus. One can embed $\cal A$ in the ``universal differential
algebra''
$\Omega^\bullet{\cal A} = \bigoplus_{n\geq 0} \Omega^n{\cal A}$,
generated by symbols
$a_0\,da_1\dots da_n$ with a formal antiderivation $d$ satisfying
$d(a_0\,da_1\dots da_n) = da_0\,da_1\dots da_n$, $d1 = 0$ and $d^2 = 0$.
Having a spectral triple allows us to condense this large algebra to a
more useful one. We first represent the whole of
$\Omega^\bullet{\cal A}$ on the Hilbert space~$\cal H$ by taking:
$$
\pi(a_0\,da_1\dots da_n) := a_0\,[D,a_1] \dots [D,a_n].
$$
The algebra of operators $\pi(\Omega^\bullet{\cal A})$ is not a
differential algebra, in general. This problem is handled by a standard
trick: the diffe\-rential ideal of ``junk''
$J := \set{c' + dc'' \in \Omega^\bullet{\cal A}: \pi c' = \pi c'' = 0}$
is factored out, thereby obtaining a new graded differential algebra of
``noncommutative differential forms'' by
$$
\Omega_D^\bullet{\cal A} := \pi(\Omega^\bullet{\cal A})/\pi(J).
$$
The quotient algebra $\Omega_D^\bullet C^\infty(M;{\Bbb C})$ for the
standard commutative spectral triple is an algebra of operators on
$L^2(S_M)$ isomorphic to the de Rham complex of differential forms.
The Connes model is given by
\begin{eqnarray*}
&& {\cal A} := C^\infty(M,{\Bbb R}) \otimes {\cal C}_F  \simeq
                   C^\infty(M,{\Bbb C}) \oplus C^\infty(M,{\Bbb H})
                   \oplus M_3(C^\infty(M,{\Bbb C})),  \\
&& {\cal H} := L^2(S_M) \otimes ({\cal H}^+_F \oplus {\cal H}^-_F),
  \qquad  D := (\delslash \otimes 1) \oplus (1 \otimes D_F).
\end{eqnarray*}

The $D_F$ operator holds information about the
Yukawa--Kobayashi--Maskawa couplings. The mimimal coupling recipe leads
then to the usual fermionic action plus the mass terms. The
noncommutative gauge potential $\Bbb A$ and field $\Bbb F$, on the boson
side, are selfadjoint elements respectively of:
\begin{eqnarray*}
\Omega_D^1{\cal A} &\simeq&
  \Lambda^1(M,{\Bbb C}) \oplus \Lambda^0(M,{\Bbb H})
   \oplus \Lambda^0(M,{\Bbb H}) \oplus \Lambda^1(M,{\Bbb H})
   \oplus M_3(\Lambda^1(M,{\Bbb C}))  \\
\Omega_D^2{\cal A} &\simeq&
  \Lambda^2(M,{\Bbb C}) \oplus \Lambda^0(M,{\Bbb H})
   \oplus \Lambda^0(M,{\Bbb H}) \oplus \Lambda^1(M,{\Bbb H})  \\
&&\qquad \oplus \Lambda^1(M,{\Bbb H}) \oplus \Lambda^2(M,{\Bbb H})
                 \oplus M_3(\Lambda^2(M,{\Bbb C})),
\end{eqnarray*}
from which the Yang--Mills Action and thus the (classical) Lagrangian
are obtained by a noncommutative procedure strictly para\-llel to the
usual one. To avoid a $U(3) \times SU(2) \times U(1)$ theory, however,
an ingredient is missing. Following Connes we impose the
``unimodularity condition''
$$
\Str ({\Bbb A} + J {\Bbb A} J) = 0,
$$
where the supertrace is taken with respect to particle-antiparticle
splitting; here $J$ is the conjugation operator that interchanges
particles and antiparticles. One gets the reduction to
the SM gauge group and the correct hypercharges; this happens now
irrespectively of whether neutrinos are massive or not.
We have recently shown that the unimodularity condition is
strictly equivalent, within the NCG framework, to anomaly cancellation:
a first exciting hint at a deeper relationship between quantum physics
and NCG than was known before.

\subsection*{Recapitulation}

The picture that emerges is that of a ``doubling'' of the space stemming
from chirality, with gauge bosons corresponding to the displacements in
continuous directions and the Higgs boson corresponding to the
exchange of quanta in the discrete direction.

There are 18 free parameters in the SM (leaving aside the vacuum angle
$\theta$): the strong coupling constant $\alpha_3$; the electroweak
para\-me\-ters $\alpha_2$, $\sin^2\theta_W$, $m_W$; the Higgs mass; the
nine (or twelve, if neutrinos are massive) fermion masses; and four
Kobayashi--Maskawa parameters. One has as inputs the fermionic constants
only; one can treat $\alpha_2$ as an adjustable parameter. When all
computations are done, one obtains the {\it constrained\/} classical SM
Lagrangian:
\begin{eqnarray*}
{\cal L} &=& - \tfrac14 A B_{\mu\nu} B^{\mu\nu}
             - \tfrac14 E F_{\mu\nu}^a F_a^{\mu\nu}
             - \tfrac14 C G_{\mu\nu}^a G_a^{\mu\nu}
             + S D_\mu\phi\, D^\mu\phi   \\
&& \qquad    - L (\|\phi_1\|^2 + \|\phi_2\|^2)^2
             + 2L (\|\phi_1\|^2 + \|\phi_2\|^2),
\end{eqnarray*}
where $B,F,G$ denote respectively the $U(1)$, $SU(2)$, $SU(3)$ gauge
fields and the coefficients $A,E,C,S,L$ are given in function of four
{\it unknown\/} parameters $C_{\ell f}$, $C_{\ell c}$, $C_{qf}$,
$C_{qc}$, which play the role of coupling constants in NCG.

The appearance of parameter restrictions is only natural, as all gauge
fields now are part of a unique field. As only the ratios among those
NCG parameters are important, there would remain only one
``prediction'', i.e., the Higgs particle mass. We can be a little more
explicit if we take the values which are more natural in the NCG
framework: $C_{\ell f} = C_{\ell c}$, $C_{qf} = C_{qc}$. Introduce the
parameter $x := (C_{\ell f} - C_{qf})/(C_{\ell f} + C_{qf})$, with
range $-1 \leq x \leq 1$. The most natural value is $x = 0.5$. When one
identifies the previous constrained Lagrangian to the usual SM
Lagrangian, it yields:
$$
m_W = \mtop \sqrt{{3 \over N_F} \, {1 - x \over 4 - 2x}}.
$$
Then $\mtop \geq \sqrt{3} m_W$. Similarly,
$g_3 = \frac12 g_2 \sqrt{(4 - 2x)/(1 - x)}$.

For the Weinberg angle, in the massive neutrino case, one gets
$\sin^2 \theta_W = (12 - 6x)/(32 - 8x)$. Then one obtains the constraint
$\sin^2 \theta_W \leq 0.45$. Finally, for the mass of the Higgs:
$$
m_H = \mtop \sqrt{3 - {3\over N_F}\,{1 - x \over 2 - x}}
= \mtop \sqrt{3 - {6 m_W^2 \over N_F \mtop^2}};
$$
from which we get the relatively tight constraint
$\sqrt{7/3}\,\mtop \leq m_H \leq \sqrt3\,\mtop$.

\subsection*{Open problems}

One can accommodate the experimental values of the strong coup\-ling
constant and the Weinberg angle by choosing $C_{\ell c} \gg C_{qc}$.
Thus, NCG offers no real predictions for the ratio of the coupling
constants to the Weinberg angle. Though $m_W \leq \mtop/\sqrt{N_F}$
is a suggestive constraint ---it gives at once the right ballpark---
there is no true prediction for the mass of the top quark, either.
Rather, the experimentally determined top mass helps to fix the more
important parameter of the theory, namely~$x$. Once the top mass is
pinned down, the model seems to fix the value of the Higgs mass. For
instance, if  $\mtop = 2.5\,m_W \simeq 200 \GeV$, we get $x = 0.53$,
and then $m_H = 328.3\,\GeV$. Note that for $x \gtrsim 0.8$, we are
outside the perturbative regime in Quantum Field Theory. If there were
a compelling reason to adopt Connes' relations on-shell, the theory
would stand or fall by the value of the Higgs mass.

On the other hand, unless and until someone comes out with a
quantization procedure specific to NCG that does the trick, there seems
to be no such compelling reason. It is only reasonable to apply the
standard renormalization procedures of present-day QFT to Connes'
version of the SM Lagrangian. The constraints are not preserved under
the renormalization flow, i.e., they do not correspond to a hidden
symmetry of the SM. The view that any constraints can be imposed only
in a fully renormalization group invariant way is, nevertheless,
theoretically  untenable.

It is just conceivable that Nature has chosen for us a scale $\mu_0$ at
which to impose Connes' restrictions. If we choose $x = 0.5$, the
present experimental values for the strong interaction coupling and
Weinberg angle are regained on imposing Connes' relations at the energy
scale $\mu_C \approx 5 \times 10^8\,\GeV$ (in the massive case). This
``intermediate unification scale'' would mark the limit of validity of
the present, phenomenological NCG model, essentially corresponding to an
ordinary, but disconnected, manifold; at higher energy scales, the
regime of truly noncommutative geometries would begin.  On imposing the
mass relations at $\mu_C$, and running the renormalization equations at
one loop, we get $\mtop \simeq 215\,\GeV$ (within the error bars of the
D0 experiment) and $m_H \simeq 235{-}240\,\GeV$. The 1-loop
approximation is not very accurate; inclusion of quantum corrections at
2nd order would give somewhat higher Higgs masses.

There is also a direct relation between NCG and gravitation: the
noncommutative integral $\ncgint D^{-2}$ gives the Einstein--Hilbert
action of general relativity. However, there seems to be at present no
unambiguous unification strand, within NCG, of gravitation and the
subatomic forces.

\subsection*{Some sources}

The original groundbreaking paper was \cite{ConnesLott}. For the
``old scheme'' of NCG (as presented in the 1992--94 period), and the
introduction of the ``new scheme'', see \cite{BookandReality}.
For the mathematics of NCG, see \cite{BookandReality} and \cite{Sirius}.
The parameter relations were derived in \cite{KastlerSchucker}.
Renormalization of NCG models, and the r\^ole of anomalies in NCG
schemes, have been explored in \cite{Orfeos}. A noncommutative geometry
model with massive neutrinos was proposed in \cite{Proteus}. Links
between gravitation and NCG have been studied in \cite{KastlerandWalze}.
For the philosophy of the whole thing, see \cite{Belindaetal}.

\end{document}